\documentclass{IEEEtaes}

\usepackage{array,amsthm}
\usepackage{amsmath,amssymb}
\usepackage{graphicx}
\usepackage{epsfig}
\usepackage{color}
\usepackage{booktabs}
\usepackage{balance}
\usepackage{mathtools}
\usepackage{algorithm}
\usepackage{algpseudocode}
\usepackage[caption=false,font=normalsize,labelfont=sf,textfont=sf]{subfig}

\jvol{XX}
\jnum{XX}
\jmonth{XXXXX}
\paper{1234567}
\pubyear{2026}
\doiinfo{TAES.2026.Doi Number}

\long\def\rchg#1{#1}

\begin{document}

\title{SCI-Mamba: Unsupervised Learning based Low-Light Image Enhancement for Non-Cooperative Spacecraft}

\author{Yiyong Sun}
\member{Member, IEEE}

\author{Weihang Shan}

\author{Shijun Wei}

\author{Diwei Zhou}

\author{Guang Zhai}
\affil{School of Aerospace Engineering, Beijing Institute of Technology, Beijing, China}



\receiveddate{This work was supported in part by the Fundamental Research Funds for the Central Universities,
the National Key Research and Development Program of China under Grant 2019YFA0706500,
National Natural Science Foundation
of China Grants (61903219), and
Beijing Institute of Technology Research
Fund Program for Young Scholars (XSQD-202101013).
}

\authoraddress{Authors' address: Yiyong Sun, Weihang Shan, Shijun Wei, Diwei Zhou, and Guang Zhai are with the School of Aerospace Engineering, Beijing Institute of Technology, Beijing, 100081, China. e-mail: (sunyy@bit.edu.cn; bitswh@bit.edu.cn; weishijun@bit.edu.cn; dvvzhou@bit.edu.cn;   gzhai@bit.edu.cn).}

\corresp{Corresponding author: Yiyong Sun (e-mail: \href{mailto:sunyy@bit.edu.cn}{sunyy@bit.edu.cn}). \\
Code, appendix and dataset: \url{https://github.com/bitswh/SCI-Mamba}}


\markboth{Y. Sun et al.}{SCI-Mamba...}
\maketitle

\begin{abstract}
\rchg{
Low-light image enhancement is a meaningful preprocessing step for on-orbit servicing of non-cooperative spacecraft, where paired training data are scarce in practice.
Among existing unsupervised approaches, CNN-based methods achieve high speed but suffer from limited receptive fields, while Transformer-based methods excel at global modeling yet incur quadratic computational and memory overhead.
Recent visual Mamba variants leverage linear-complexity state-space modeling to balance efficiency and accuracy, but they still require repeated layer-wise 2D/1D conversions introducing redundant cross-dimensional overhead.
To address this limitation, this paper proposes SCI-Mamba, an unsupervised low-light enhancement network that adopts a fully sequence-dominated design: all illumination estimation and refinement operations are executed in 1D sequence space, with only one global flattening at input and a single reconstruction at output.
The framework further integrates self-calibrated illumination decomposition, Retinex physical priors, and a prior-guided illumination monitor enabling spatially adaptive enhancement for spacecraft foreground and deep-space background.
We construct \textit{Space Dark-1.0}, a multi-source dataset combining synthetic renders, darkroom hardware-in-the-loop captures, and real on-orbit footage.
Experiments show SCI-Mamba runs $20.53\times$ faster than ECMamba and $4.40\times$ faster than WalMaFa with 394.58 MB peak memory, delivering $3.44\times$--$9.62\times$ speedup over evaluated Transformer and flow-based methods while producing visually natural results for visual inspection.
}
\end{abstract}


\section{Introduction} \label{Sec::Introduction}
Low-light image enhancement is a \rchg{meaningful} preprocessing step for visual perception in non-cooperative spacecraft on-orbit servicing, \rchg{since it can affect the accuracy of} downstream tasks \rchg{such as} autonomous rendezvous \cite{velentzas2025irolsd}, pose estimation \cite{peng2019pose}, key component detection \cite{khalil2025hybrid}, and robotic capture \cite{cao2024spacecraft,xu2011survey}\rchg{, and improves the interpretability of inspection imagery for ground operator monitoring and post-mission analysis}.
Images captured under orbital \rchg{low-light} conditions suffer from composite degradation consisting of \rchg{substantial} sensor noise, local over-saturation, and motion blur\rchg{, which further degrades both machine vision accuracy and human visual perception quality}.
\rchg{Paired} low-light and normal-light images of spacecraft are \rchg{rare in practice, which makes it} difficult for \rchg{supervised methods} to generalize.
\rchg{Developing} paired-data-free unsupervised enhancement algorithms \cite{zhao2025review,ma2022toward,ma2025learning} tailored for space imaging is therefore \rchg{of practical value}.

Traditional low-light enhancement methods built on hand-crafted priors mainly include \rchg{histogram equalization \cite{zuiderveld1994contrast}, Gamma correction \cite{huang2012efficient}, and Retinex decomposition \cite{land1977retinex,guo2016lime}}.
When deployed for dynamic, high-contrast orbital scenes with non-cooperative targets, these approaches suffer from weak scene adaptability, over-amplified background noise, and severe color artifacts.
Histogram equalization boosts global contrast via gray-scale mapping but simultaneously magnifies sensor noise and induces highlight overexposure, losing tiny structural details.
Gamma correction adjusts overall brightness through non-linear transformation yet fails to cope with complex, spatially varying orbital illumination, resulting in limited restoration gains.
\rchg{Retinex-based methods offer clear physical interpretability by separating illumination and reflectance, yet their manually predefined regularization terms (e.g., illumination smoothness and reflectance sparsity) degrade robustness under dynamic space lighting \cite{an2025striving,cai2023retinexformer,bai2024retinexmamba}, and division-based reflectance recovery under low light amplifies noise, causing color shifts and halo artifacts.}

With the development of deep learning, CNN-based unsupervised methods have gained traction in low-light enhancement.
\rchg{The Self-Calibrated Illumination (SCI)} framework introduces a learning module to implement progressive illumination-reflectance decomposition with fast inference speed, and its upgraded variant SCI++ further strengthens model robustness via multi-stage self-calibration \cite{ma2022toward,ma2025learning}.
Zero-DCE++ \cite{li2021learning} formulates enhancement as high-order brightness curve estimation without requiring paired data.
All of the above CNN-based methods eliminate dependence on matched light-dark image pairs and meet the data constraints of space applications.
Nevertheless, the fixed local receptive field of convolutional layers restricts global context modeling, leading to uneven restoration effects in spacecraft scenes with extreme brightness gaps between vehicle bodies and dark cosmic backgrounds \cite{wang2024zero}.

To overcome CNNs' long-range modeling limitation,
Transformer-based methods \rchg{naturally possess global context modeling via self-attention} \cite{feng2026dmat}. Uformer \cite{wang2022uformer} adopts a U-shaped architecture with locally enhanced windows \rchg{for image restoration}, LLFormer \cite{wang2023ultra} designs attention-based brightness enhancement and denoising modules for ultra-high-definition low-light images, and UHDformer \cite{wang2024correlation} introduces correlation matching transformation for image enhancement. These attention-based models dynamically redistribute feature weights according to regional brightness\rchg{,} matching \rchg{the} high contrast of space target imagery \cite{cai2023retinexformer,wang2024division}.
\rchg{However, the quadratic computational complexity of self-attention creates high computation and memory costs that exceed} spaceborne payload limits, \rchg{and even lightweight or sparse variants cannot eliminate this overhead for} high-resolution feature sequences \cite{bai2024retinexmamba}.

\rchg{Recently, the linear-complexity Mamba state-space model (SSM) has shown unique advantages in efficient sequence modeling, remote sensing and radar} \cite{xu2025smnet,wang2026advstmamba}\rchg{, delivering competitive performance with lower cost; its selective mechanism acts as an adaptive filter} \cite{gu2023mamba}\rchg{ capturing long-range dependencies} \cite{liu2024vmamba}. VMamba \cite{liu2024vmamba} introduces a 2D selective scan (SS2D) \rchg{that unfolds images into four directional 1D sequences via cross-scanning and applies a selective SSM with linear complexity, avoiding the quadratic cost of self-attention}. In low-light enhancement, \mbox{ECMamba} \cite{dong2024ecmamba} combines Retinex guidance with Mamba for \rchg{exposure correction}, while WalMaFa \cite{tan2024wavelet} incorporates wavelet and \rchg{frequency-domain adjustment}.
However, several limitations persist when applied to space target scenarios: existing visual Mamba architectures, like VMamba \cite{liu2024vmamba}, require repeated \rchg{conversions between 2D feature maps and 1D sequences, which introduce additional computational overhead and may affect the preservation of spatial structures}.
These methods are primarily designed for ground scenes and have not been validated in space imaging scenarios; \rchg{whether they can enhance spacecraft targets and deep-space backgrounds differently} is still an open question \cite{park2023adaptive,wang2024division}.

Apart from algorithmic limitations, dedicated benchmark datasets for space low-light enhancement remain absent. Widely used orbital target datasets such as SISIFOS \cite{velentzas2026sisifos}, SpaceSeg \cite{liu2025spaceseg}, SPEED \cite{kisantal2020satellite}, and SPEED+ \cite{park2022speed+} are designed \rchg{for pose estimation}, consisting mostly of generic synthetic samples without authentic low-light orbital measurements. Spacecraft-DS \cite{cao2024spacecraft} focuses on \rchg{component detection}, without covering typical low-light orbital operating environments. \rchg{Ray-tracing-based} datasets \cite{bechini2023dataset} can generate high-fidelity imagery, but due to the absence of hardware-in-the-loop capture, they exhibit \rchg{domain gaps} from actual spaceborne cameras in terms of sensor noise and dynamic range \cite{proencca2020deep}.

To address the above challenges, this paper integrates the efficient unsupervised SCI++ framework \cite{ma2025learning} with the VMamba architecture \cite{liu2024vmamba}, incorporates Retinex-based physical priors \cite{an2025striving}, and proposes SCI-Mamba, an unsupervised low-light enhancement method specifically tailored for non-cooperative spacecraft.
We design a fully sequence-dominated pipeline that executes exactly one 2D-to-1D flattening at the input and one 1D-to-2D reconstruction at the output; all core illumination estimation and self-calibration procedures operate purely within 1D sequence space to eliminate redundant cross-dimensional transformation overhead.
A multi-objective joint loss suite is further proposed, equipped with a prior-guided illumination monitor that automatically generates target illumination maps from input brightness and gradient features to realize differentiated foreground-background enhancement.
Furthermore, a dedicated low-light dataset is constructed, encompassing physically constrained synthetic images, hardware-in-the-loop darkroom images, and real on-orbit imagery.
The main contributions of this paper are summarized as follows:
\begin{itemize}
   \item A sequence-inference-dominated enhancement architecture, SCI-Mamba, is proposed. The pipeline restricts cross-dimensional transformation to one forward and one reverse conversion only, avoiding redundant feature computation and boosting inference efficiency. Compared with \rchg{a 2D-scan ablation variant}, the average inference frame rate improves by \rchg{19.9\%}.
    \item A dedicated space low-light dataset, \textit{Space Dark-1.0}, is constructed. It contains physically constrained software-synthesized images, darkroom hardware-in-the-loop satellite model images, and real on-orbit images, filling the gap in specialized space low-light datasets.
    \item A multi-objective loss function containing color fidelity and smoothness denoising is designed, in which the prior-guided illumination monitor module automatically generates a target illumination map from the brightness and gradient of the input image, guiding the network to focus on texture-rich regions for differentiated enhancement.
    \item \rchg{Comparative} inference experiments conducted on the \textit{Space Dark-1.0} test set verify the adaptive enhancement capability of SCI-Mamba.
        Unlike CNN-based unsupervised competitors that tend to amplify noise indiscriminately in textureless background regions, our method achieves a balanced trade-off between targeted foreground enhancement and effective background noise suppression for orbital scenes.
        In terms of inference speed, SCI-Mamba runs \rchg{$20.53\times$ and $4.40\times$} faster than ECMamba and WalMaFa, respectively.
        Compared with LLFlow, Uformer, LLFormer, and UHDformer, SCI-Mamba runs \rchg{$3.44\times$, $8.17\times$, $9.62\times$, and $3.68\times$} faster, \rchg{respectively}.

\end{itemize}

The remainder of this paper is organized as follows:
Section~\ref{Sec::MainResult} elaborates on the proposed SCI-Mamba method, including the overall framework, the sequence-dominated cross-scan enhancement mechanism, and \rchg{the 1D Visual State Space module (VSS1D).}
Section~\ref{Sec::Loss_Function} introduces the multi-objective loss function, including the prior-guided illumination monitor.
Section~\ref{Sec::Space_Dark_and_Training} describes the \textit{Space Dark-1.0} dataset and the training process.
Section~\ref{Sec::Experiment} presents quantitative and qualitative
experimental comparisons against mainstream restoration algorithms.
Section~\ref{Sec::Conclusion} concludes the paper and outlines future research directions.

\section{SCI-Mamba} \label{Sec::MainResult}

\subsection{Overview}\label{SubSec::GeneralDescription}
\begin{figure*}[htp]
	\centering
	\includegraphics[width=0.9\linewidth]{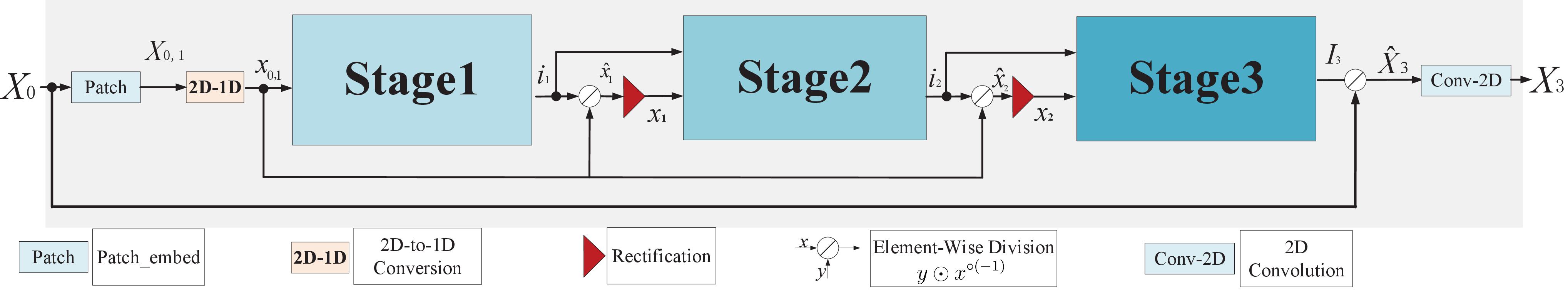}
	\caption{Structure of SCI-Mamba}
	\label{Fig::SCI_Mamba_Structure}
\end{figure*}

The overall SCI-Mamba architecture is visualized in Figure~\ref{Fig::SCI_Mamba_Structure}.
This framework integrates the self-calibrated enhancement mechanism of SCI++ with the efficient sequence modeling capability of VMamba, establishing a sequence-dominated enhancement architecture that fundamentally eliminates repeated cross-dimensional transformation.
The layer-wise conversion overhead of conventional Mamba variants, discussed in the introduction, is \rchg{removed by flattening the image to a 1D sequence only once, making} the pipeline adaptable to resource-limited spaceborne \rchg{hardware}.

As shown in Figure~\ref{Fig::SCI_Mamba_Structure}, \rchg{the input image is first resized to $X_0^l$ when its longer side exceeds 512 pixels (Algorithm~\ref{alg:sci_mamba_flow});} \rchg{the resized image then} passes through a patch embedding layer to generate a downsampled low-resolution feature map $X_{0,1}$.
A single global 2D-to-1D flattening operation converts $X_{0,1}$ into the 1D backbone sequence $x_{0,1}$ for all subsequent computations.
Only at the final reconstruction stage \rchg{is a single 1D-to-2D reshaping executed; this single reconstruction avoids} the layer-wise domain switching overhead \rchg{of} existing visual Mamba variants.

The network consists of three progressive illumination refinement stages, all performing feature propagation and optimization entirely in the sequence domain: \textbf{Stage 1 (Initial Illumination Estimation)}: \rchg{It} receives the initial sequence $x_{0,1}$ and generates an initial illumination sequence $i_1$; sequence-domain Retinex enhancement is then applied to produce the initial enhanced sequence $\hat{x}_1$, which is subsequently rectified via numerical truncation to yield $x_1$. \textbf{Stage 2 (Illumination Correction)}: \rchg{It} introduces a self-calibration mechanism that performs long-range dependency modeling on the sequence $x_1$ output from Stage 1, generating a refined illumination estimation sequence $i_2$; sequence-domain Retinex enhancement is again performed using the original sequence $x_{0,1}$ as the basis, producing the corrected sequence $x_2$. \textbf{Stage 3 (Illumination Correction and Reconstruction)}: \rchg{It} generates the ultimate illumination sequence $i_3$ based on the sequence $x_2$; image-domain Retinex enhancement is subsequently applied to produce $X_3$, followed by a final 2D convolution that yields the ultimate output.

The three stages are not merely cascaded enhancements but \rchg{center} on the progressive refinement of illumination estimation, \rchg{successively improving} dark-region enhancement and detail recovery while maintaining the physical constraints \rchg{of} Retinex theory.

\rchg{Retinex theory is used throughout the three enhancement stages as the physical basis for illumination-based reconstruction.}
The standard image-domain Retinex recovery formulation is defined as \cite{ma2025learning,cai2023retinexformer,yi2025diff}
\begin{equation}
\hat{X}_3 = X_0 \odot I_3^{\circ(-1)} \label{Eq::Retinex_Image}
\end{equation}
where $I_3^{\circ(-1)}$ denotes the element-wise Hadamard inverse of the illumination map $I_3$, and $\odot$ represents element-wise Hadamard multiplication.
\rchg{The 2D convolutional module following $\hat{X}_3$ in Figure~\ref{Fig::SCI_Mamba_Structure} is used for final image refinement. It performs local feature processing using a $3\times3$ spatial convolution with a stride of 1. A one-pixel reflection padding is applied before the convolution to preserve the spatial dimensions, while replication padding is used when the spatial resolution is too small for reflection padding. The residual correction predicted by this module is added to the range-clipped reconstruction to obtain the final enhanced image $X_3$. The convolutional parameters are jointly optimized during network training.}
To adapt Retinex decomposition to the sequence-dominated pipeline, we extend the transformation into 1D sequence space for Stage 1 and Stage 2 processing
\begin{equation}
\hat{x}_1 = x_{0,1}\odot i_1^{\circ(-1)}, \quad
\hat{x}_2 = x_{0,1}\odot i_2^{\circ(-1)}
\label{Eq::Retinex_Sequence}
\end{equation}
All Retinex operations for intermediate refinement are completed in the 1D domain without frequent image-sequence switching, which \rchg{reduces repeated image-sequence conversions and contributes to} the computational efficiency of the pipeline.

\subsection{Stage-Wise Enhancement Pipeline} \label{SubSec::Stage_wise_Enhancement}

\subsubsection{Stage 1: Initial Illumination Estimation}
\begin{figure}[htpb]
	\centering
		\includegraphics[width=1\linewidth]{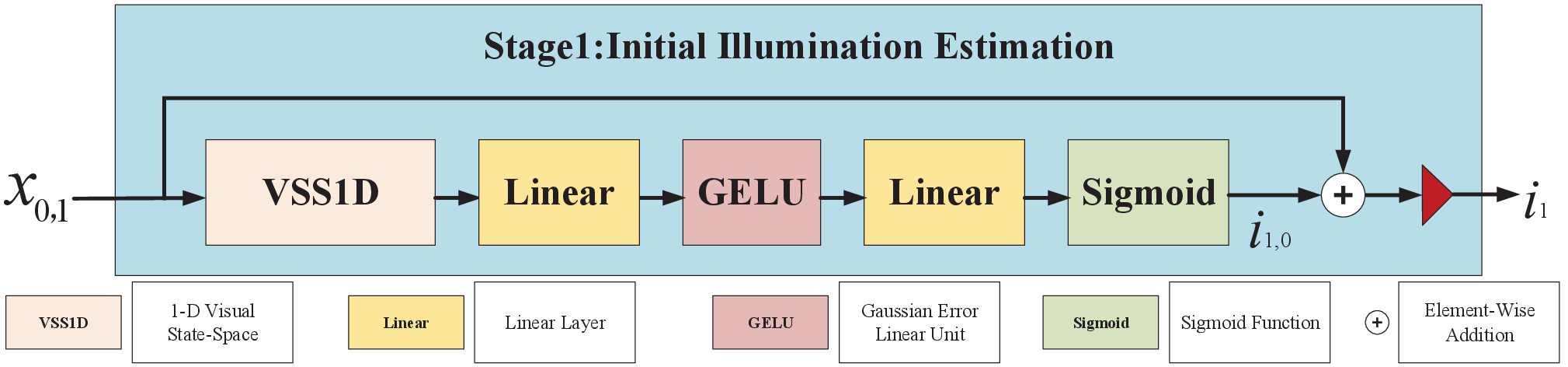}
	\caption{Structure of Stage 1}
	\label{Fig::SCI_Mamba_Stage1}
\end{figure}

The architecture of Stage 1 is shown in Figure~\ref{Fig::SCI_Mamba_Stage1}. This module undertakes coarse global illumination estimation from the flattened input sequence $x_{0,1}$, relying on the proposed VSS1D module (detailed in Section~\ref{Subsubsec::VSS1D}) to extract long-range and local spatial features from low-light sequence data.

After VSS1D feature extraction, linear projection and Sigmoid activation generate raw illumination compensation sequence $i_{1,0}$. Element-wise summation with original $x_{0,1}$ and $[0,1]$ clipping produces the initial illumination estimation $i_1$, which integrates original brightness information with global contextual compensation modeled by VSS1D. We then execute sequence-domain Retinex enhancement following \eqref{Eq::Retinex_Sequence}, and rectify the output value range to get Stage 1 intermediate sequence $x_1$ fed into Stage 2.

\subsubsection{Stage 2: Illumination Correction }
\begin{figure}[htpb]
	\centering
		\includegraphics[width=1\linewidth]{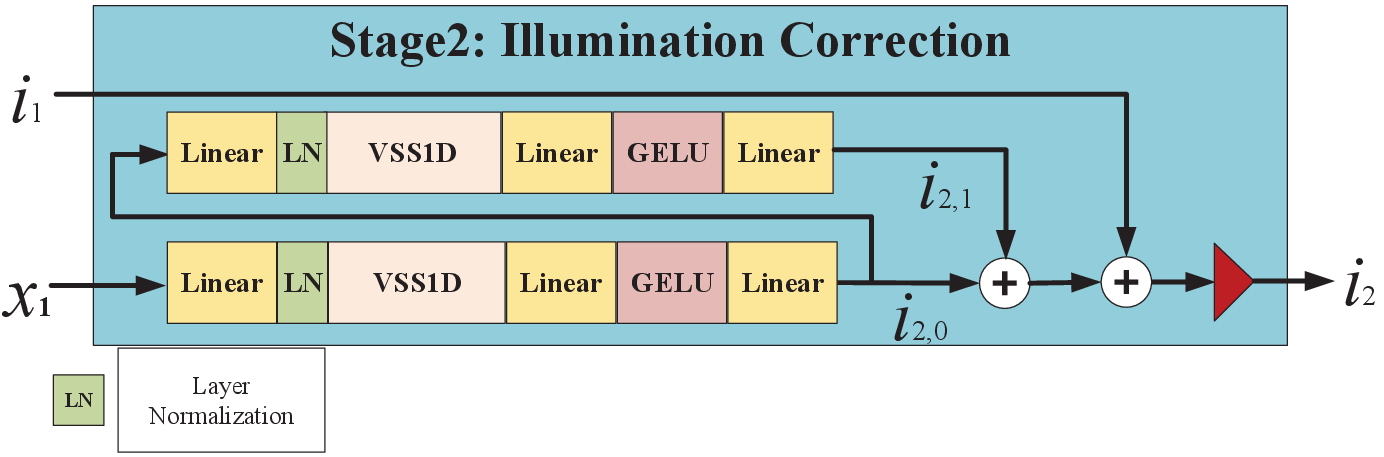}
	\caption{Structure of Stage 2}
	\label{Fig::SCI_Mamba_Stage2}
\end{figure}

Figure~\ref{Fig::SCI_Mamba_Stage2} illustrates the self-calibration refinement module of Stage 2. Distinct from Stage 1's single-input illumination generation logic, this stage accepts both the enhanced intermediate $x_1$ and illumination estimation $i_1$ from the prior stage, inheriting the multi-stage paradigm of SCI++ \cite{ma2025learning}.

The input $x_1$ passes through layer normalization, linear projection and two stacked VSS1D units to derive two hierarchical illumination correction terms $i_{2,0}, i_{2,1}$. Summation of these two terms with Stage 1's $i_1$, followed by value clipping, yields refined illumination sequence $i_2$.
To avoid progressive error accumulation across cascaded stages, Retinex enhancement is always computed against the original base sequence $x_{0,1}$ rather than refined $x_1$. The resulting $\hat{x}_2$ undergoes range rectification to generate $x_2$, the input feature sequence for Stage 3 reconstruction.
This design prevents error propagation from Stage 1 intermediate outputs into subsequent illumination estimation.

\subsubsection{Stage 3: Illumination Correction and Reconstruction}
\begin{figure}[htpb]
	\centering
		\includegraphics[width=1\linewidth]{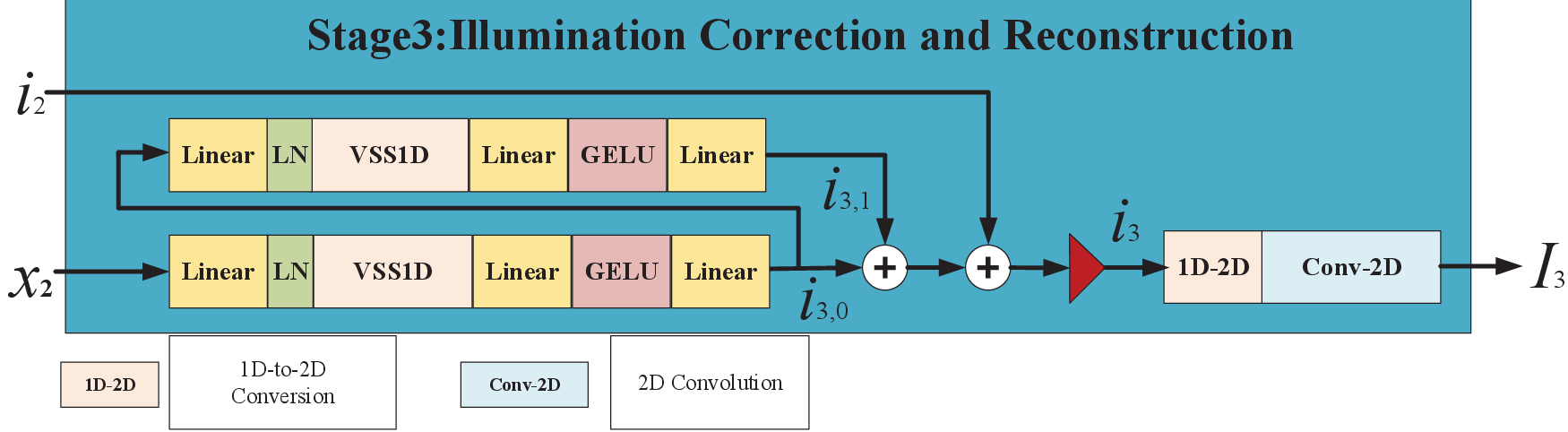}
	\caption{Structure of Stage 3}
	\label{Fig::SCI_Mamba_Stage3}
\end{figure}

The final reconstruction module is displayed in Figure~\ref{Fig::SCI_Mamba_Stage3}. Its illumination refinement sub-process shares identical VSS1D-based optimization logic with Stage 2, taking $x_2$ to compute the ultimate 1D illumination sequence $i_3$ after truncation. The key functional difference lies in the additional cross-dimensional transformation reconstruction branch unique to Stage 3.

The 1D sequence $i_3$ is reshaped into a 2D patch grid via the pipeline's sole 1D-to-2D cross-dimensional transformation operation. \rchg{Specifically, the final illumination sequence $i_3$ has a size of $B\times(h_pw_p)\times3$, where $B$ denotes the batch size, and $h_p$ and $w_p$ denote the two-dimensional grid dimensions preserved during input embedding. The sequence and channel dimensions are first permuted, after which the sequence is rearranged in row-major order into a two-dimensional illumination grid $G_3\in\mathbb{R}^{B\times3\times h_p\times w_p}$. This rearrangement only restores the two-dimensional spatial layout and introduces neither learnable parameters nor interpolation. The spatial resolution is subsequently recovered by the following upsampling operation.
}\
Successive upsampling and 2D convolution recover full-resolution illumination map $I_3$ aligned with input $X_0$.
\rchg{\
The illumination map is reconstructed using bilinear interpolation followed by channel-wise convolutional refinement with the same $3\times3$ reflection-padded convolution described above.
}\
 Image-domain Retinex transformation\eqref{Eq::Retinex_Image} is applied on raw input $X_0$ using $I_3$, and a terminal convolutional layer generates the final enhanced output image $X_3$.

\subsection{Sequential State-Space Modeling in 1D}
Standard VMamba \cite{liu2024vmamba} and Transformer-based restoration networks,  such as Uformer \cite{wang2022uformer}, LLFormer \cite{wang2023ultra}, and UHDformer \cite{wang2024correlation}, operate on 2D feature maps, requiring layer-wise 2D-to-1D flattening and inverse reshaping for SSM/self-attention computation.
Repeated cross-dimensional transformation between image grids and 1D sequences introduces heavy redundant computation and breaks continuous spatial feature representation, limiting real-time inference speed for low-power hardware. \rchg{VMamba's VSS blocks perform a 2D-to-1D flattening and an inverse reshaping in every layer, incurring $2M$ cross-dimensional conversions for an $M$-layer backbone. In contrast, SCI-Mamba performs exactly two such conversions, one at the input and one at the output, so the conversion overhead is reduced from $\mathcal{O}(M)$ to $\mathcal{O}(1)$. The four-directional scanning scheme is inherited from VMamba/SS2D; our contribution lies in embedding it in a fully sequence-based architecture that removes the layer-wise domain switching.}
\rchg{\
To reduce this repeated conversion overhead, we design the VSS1D module family including Sequential Cross-Scan Generator (SCSG) and Sequential Cross-Scan Restorer (SCSR) sub-units.
}\
SCSG generates four directional scan sequences within 1D tensors, while SCSR fuses multi-path features back to row-major order without reconstructing 2D feature maps.

\subsubsection{VSS1D} \label{Subsubsec::VSS1D}
\begin{figure}[htpb]
	\centering
		\includegraphics[width=1\linewidth]{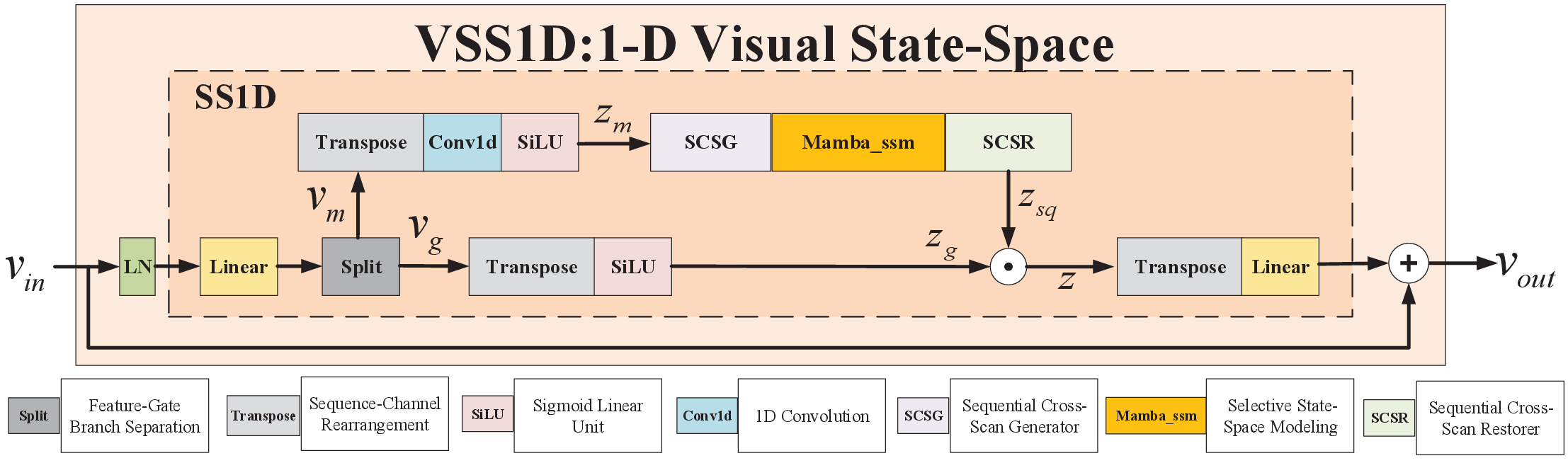}
	\caption{Structure of VSS1D}
	\label{Fig::SCI_Mamba_Structure_5}
\end{figure}

We redesign the original VMamba VSS block into VSS1D, a residual-form module customized for flattened 1D patch sequences, as visualized in Figure~\ref{Fig::SCI_Mamba_Structure_5}. Unlike the original 2D VSS that processes grid-shaped feature maps and triggers repeated layer-wise cross-dimensional transformation, VSS1D receives the single global flattened sequence $x_{0,1}$ after patch embedding, eliminating repeated domain switching between layers.

Let the input sequence be \rchg{$v_{in}\in\mathbb{R}^{B\times L\times C}$, where $B$ denotes the batch size, $L$ the sequence length, and $C$ the channel dimension}. The forward propagation formula of VSS1D is
\begin{equation}
v_{out} = v_{in} + \operatorname{SS1D} \left( \operatorname{LN}(v_{in}) \right),
\label{Eq::VSS1D}
\end{equation}
where $\operatorname{LN}(\cdot)$ denotes layer normalization (LN), $\operatorname{SS1D}(\cdot)$ is the embedded 1D selective scan (SS1D) core, and \rchg{$v_{out}\in\mathbb{R}^{B\times L\times C}$} denotes module output. The residual shortcut retains raw sequence features to mitigate gradient instability in deep multi-stage propagation.

SS1D serves as the core spatial modeling sub-block inside VSS1D. After layer normalization, input features are linearly expanded and split into two parallel branches: spatial state modeling branch $v_m$ and adaptive gating branch $v_g$
\begin{equation}
v_{m},v_{g} = \operatorname{Split} \left( \operatorname{Linear} \left( \operatorname{LN}(v_{in}) \right) \right). \label{Eq::SS1D_Split}
\end{equation}
The spatial branch first applies transposition, 1D convolution and SiLU activation to strengthen local adjacent pixel correlations in the flattened sequence
\begin{equation}
{z}_{m} = \operatorname{SiLU} \left( \operatorname{Conv1D} \left( \operatorname{Transpose}(v_{m}) \right) \right). \label{Eq::SS1D_Conv}
\end{equation}
A single-direction scanning sequence fails to capture horizontal, vertical, and reverse spatial relationships inherited from the original 2D patch grid. We therefore embed the SCSG generator inside the spatial branch to produce four directional scan sequences purely from 1D data, detailed in Section~\ref{Subsubsec::SCSG}.

Each directional sequence from SCSG is fed into Mamba\_ssm selective state-space modeling to capture long-range cross-patch dependencies under independent propagation paths. After SSM computation, the SCSR restoration module (Section~\ref{Subsubsec::SCSR}) unifies all directional feature sequences back to the original row-major order and executes balanced feature fusion
\begin{equation}
{z}_{sq} = \operatorname{SCSR} \left( \operatorname{Mamba\_ssm} \left( \operatorname{SCSG} \left({z}_{m} \right) \right) \right).
\label{Eq::SS1D_Scan}
\end{equation}
Simultaneously, the gating branch processes \rchg{the} split $v_g$ via \rchg{a} transposition and SiLU to generate adaptive modulation weights
\begin{equation}
{z}_{g} = \operatorname{SiLU} \left( \operatorname{Transpose}({v}_{g}) \right).
\label{Eq::SS1D_Gate}
\end{equation}
\rchg{Element-wise multiplication between spatial modeling output $z_{sq}$ and gating response $z_g$  suppresses useless feature responses and emphasizes structure-related features}
\begin{equation}
{z} = {z}_{sq} \odot {z}_{g}.
\label{Eq::SS1D_GatedFusion}
\end{equation}
Finally, transposed linear projection recovers the original feature dimension for SS1D output
\begin{equation}
\operatorname{SS1D}(\operatorname{LN}(v_{in})) = \operatorname{Linear} \left( \operatorname{Transpose}({z}) \right). \label{Eq::SS1D_OutProj}
\end{equation}
This SS1D output is added back to residual input $v_{in}$ to produce final VSS1D sequence $v_{out}$.

\rchg{VSS1D thus keeps} Mamba's linear-complexity long-sequence modeling \rchg{while restoring the multi-directional spatial context lost during flattening through} the embedded SCSG/SCSR units. It \rchg{supplies spatially consistent} sequence representations for all three illumination\rchg{-refinement} stages.

\subsubsection{From SS2D to SS1D Sequence Modeling} \label{Subsubsec::SS1D}
\begin{figure}[htbp]
	\centering
		\includegraphics[width=1\linewidth]{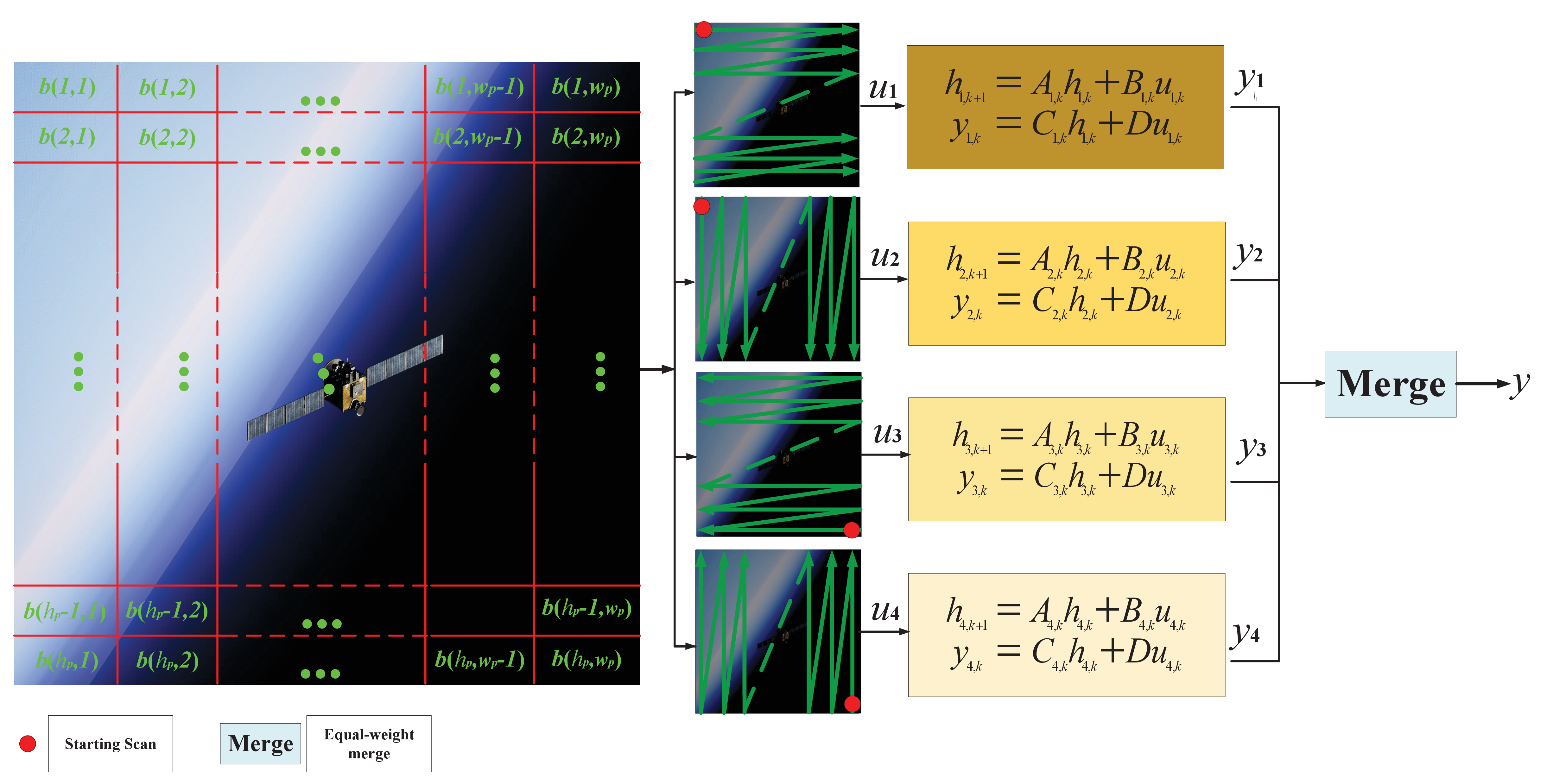}
	\caption{Illustration of the 1D Sequential Cross-Scan Mechanism}
	\label{Fig::SS2DMechinism}
\end{figure}

VMamba's native SS2D module constructs four directional scan sequences from intact 2D feature maps, capturing multi-orientation spatial correlations via row/column forward/reverse traversal, as illustrated in Figure~\ref{Fig::SS2DMechinism}.

\rchg{Denote $b(i,j)$ as the feature vector at row $i$, column $j$ of the patch grid, with grid dimensions $h_p$ (height) and $w_p$ (width).} SS2D generates four complementary traversal sequences
\begin{equation}
\begin{array}{rl}
u_1 : & b(1,1), b(1,2),..., b(1,w_p), b(2,1),...,b(2,w_p),\\
      & ...,b(h_p,1), ...,b(h_p,w_p) \\
u_2 : & b(1,1), b(2,1),..., b(h_p,1), b(1,2),...,b(h_p,2),\\
      &...,b(1,w_p), ...,b(h_p,w_p) \\
u_3 : & b(h_p,w_p), b(h_p,w_p-1),..., b(h_p,1),  b(h_p-1,w_p),\\
      &...,b(h_p-1,1), ...,b(1,w_p),...,b(1,1) \\
u_4 : & b(h_p,w_p), b(h_p-1,w_p),..., b(1,w_p), b(h_p,w_p-1),\\
      &...,b(1,w_p-1), ...,b(h_p,1), ...,b(1,1) \\
\end{array}.
\label{Equ::SS2D_Mechanism}
\end{equation}
Each sequence captures complementary horizontal/vertical/reverse contextual cues to overcome single-direction modeling bias. As analyzed in Section \ref{Sec::MainResult}, conventional SS2D introduces cumulative transformation overhead when stacked; it requires every module to receive 2D grid inputs and execute internal flattening/recovery operations.
\rchg{SCI-Mamba} pipeline flattens the image once at input;
\rchg{reconstructing 2D maps during intermediate stages would reintroduce repeated reshaping operations that the proposed architecture aims to avoid.}
We therefore design SS1D, which generates, models and fuses four directional sequences entirely inside 1D data without temporary 2D grid recovery. The detailed generation and restoration logic are separated into SCSG and SCSR submodules.

\subsubsection{Sequential Cross-Scan Generator: SCSG} \label{Subsubsec::SCSG}
Although $z_m$ is formatted as a single flattened 1D vector, its internal pixel order strictly follows row-major patch grid arrangement. Direct single-path SSM scanning cannot extract vertical and reverse structural dependencies, so SCSG rearranges the base sequence into four independent directional traversal streams within the 1D domain.
Define $\pi_{sp}(\cdot)$ as the row-major to column-major permutation operation compatible with patch grid dimensions, and $\operatorname{Rev}(\cdot)$ as full sequence reversal. The four scan sequences generated by SCSG are
\begin{equation}
\begin{aligned}
{u}_1 &= z_{m}, \quad {u}_2 = \pi_{sp}(z_{m}),\\
{u}_3 &= \rchg{\operatorname{Rev}(z_{m})}, \quad {u}_4 = \rchg{\operatorname{Rev}\left(\pi_{sp}(z_{m})\right)}.
\end{aligned}
\label{Eq::SCSG_Four_Directions}
\end{equation}
$u_1, u_2, u_3, u_4$ correspond to row-forward, column-forward, row-reverse, and column-reverse traversal respectively. SCSG stacks the four streams along a temporary direction dimension and merges it into batch dimension for parallel batched Mamba SSM computation; this rearrangement only modifies tensor axis organization without altering feature values or introducing extra trainable weights. It supplements vertical and reverse spatial context for the flattened sequence while maintaining the 1D computation paradigm.

\subsubsection{Selective State-Space Modeling} \label{Subsubsec::Mamba}
Four directional sequences output by SCSG are processed in parallel by the Mamba\_ssm selective state-space unit. The discrete state transition dynamics for each stream $i$ at step $k$ are defined as
\begin{equation}
\left\{
\begin{aligned}
h_{i,k+1} &=& A_{i,k} h_{i,k} + B_{i,k} u_{i,k}\\
y_{i,k}   &=& C_{i,k} h_{i,k} + D u_{i,k}
\end{aligned},
\right.
 \operatorname{for}  i=1,..., 4 \label{Eq::Mamba_SSM}
\end{equation}
where \(h_{i,k}\in\mathbb{R}^N\) denotes the hidden state vector of the \(i\)-th sequence at step \(k\), \rchg{\(u_{i,k}\in\mathbb{R}^{C}\) is the input feature vector at step \(k\) of the \(i\)-th sequence \(u_i\), \(C\) denotes the channel dimension, and \(y_{i,k}\in\mathbb{R}^{C}\) is the output feature vector at step \(k\) of the \(i\)-th sequence}. \(D\) is optional and typically set to 0.
\(\bar{A}=\operatorname{diag}(-1,-2,\dots,-N)\in\mathbb{R}^{N\times N}\) is a fixedly initialized diagonal matrix in the continuous domain.
The discretized state transition matrix \(A_{i,k}\) is
\begin{equation}
A_{i,k}=\exp\left(\Delta_{i,k}\cdot\bar{A}\right)\in\mathbb{R}^{N\times N}
\end{equation}
where \(\exp(\cdot)\) denotes the exponential operation applied element-wise to matrices or scalars.
The dynamic parameters \(B_{i,k}\), \(C_{i,k}\), and \(\Delta_{i,k}\) are uniformly generated from the input via shared linear layers
\begin{equation}
\begin{array}{rcl}
B_{i,k} &=& W_B \cdot u_{i,k}  ,
C_{i,k}  = W_C \cdot u_{i,k}  , \\
\Delta_{i,k} &=& W_{dt} \cdot \left( W_{\Delta} \cdot u_{i,k} \right) + b_{\Delta}.
\end{array}
\end{equation}
The time step coefficient \(\Delta_{i,k}\in\mathbb{R}\) is passed through the \(\operatorname{Softplus}\) function with range constraints to obtain a stable positive time step
\begin{equation}
\Delta_{i,k}=\operatorname{Clamp}\left(\operatorname{Softplus}\left(\Delta_{i,k}\right),\Delta_{\min},\Delta_{\max}\right)
\label{Eq::Mamba_Delta}
\end{equation}
where \(\operatorname{Softplus}(\cdot)\) ensures positivity, defined as \(\operatorname{Softplus}(x)=\log\left(1+\exp(x)\right)\). \(\Delta_{\min}\) and \(\Delta_{\max}\) denote the lower and upper bounds of the time step coefficient, respectively.
Enforcing positivity and range constraints prevents \rchg{too small or too large state update steps, which would impede context propagation or destabilize transitions, thereby enhancing numerical stability.} \rchg{All four sequences share the same linear-layer parameters; owing to} different inputs \(u_{i,k}\), \rchg{their} dynamic parameters \(B_{i,k}\), \(C_{i,k}\), and \(\Delta_{i,k}\) \rchg{remain} mutually independent.

In summary, after the four-direction cross-scan sequences undergo Mamba selective state-space modeling as in \eqref{Eq::Mamba_SSM}, the corresponding four groups of output features are obtained
\begin{equation}
{y}_i=\operatorname{Mamba\_ssm}\left({u}_i\right),
\quad i=1,\ldots,4
\label{Eq::Mamba_Four_Directions_Output}
\end{equation}
where \({y}_1\), \({y}_2\), \({y}_3\), and \({y}_4\) represent the row-forward, column-forward, row-reverse, and column-reverse feature sequences, respectively. These four responses remain in their respective scanning orders and cannot be directly fused; they must be unified under a common encoding sequence. In this work, the SCSR module restores all four sequences to \rchg{the row-forward form}. The directional recovery and fusion process is detailed in Section~\ref{Subsubsec::SCSR}.

The Mamba module establishes long-range dependencies along different scanning directions, enabling information exchange between distant sequence elements through state recurrence. \rchg{For low-light spacecraft images, this mechanism captures contextual relationships among local structures, deep-space backgrounds, and regions of varying illumination} while maintaining linear complexity, thereby providing global structural information for subsequent illumination correction and reflectance recovery.

\subsubsection{Sequential Cross-Scan Restorer: SCSR} \label{Subsubsec::SCSR}
After Mamba SSM processing, four directional outputs $y_1,y_2,y_3,y_4$ follow inconsistent pixel permutation orders, which will break spatial alignment if fused directly. SCSR reverses SCSG's permutation operations and unifies all streams back to the original row-major sequence layout before balanced feature fusion.
Row-direction features are combined by summing forward row stream $y_1$ and reversed row-reverse stream $\operatorname{Rev}(y_3)$
\begin{equation}
{y}_{\mathrm{row}}={y}_1+\operatorname{Rev}\left({y}_3\right)
\label{Eq::SCSR_Row}
\end{equation}
Column-direction temporary features integrate column-forward $y_2$ and reversed column-reverse $y_4$
\begin{equation}
{y}_{\mathrm{col}}'={y}_2+\operatorname{Rev}\left({y}_4\right)
\label{Eq::SCSR_Col_Temp}
\end{equation}
$y_{\mathrm{col}}'$ remains column-major ordered, so we apply inverse permutation $\pi_{sp}^{-1}$ to restore row-major layout
\begin{equation}
{y}_{\mathrm{col}}=\pi_{sp}^{-1}\left({y}_{\mathrm{col}}'\right).
\label{Eq::SCSR_Col_Restore}
\end{equation}
\rchg{The four directional streams are finally merged into a unified row-major fused sequence}
\begin{equation}
\begin{aligned}
{z}_{sq}
& = \operatorname{Merge}\left({y}_1, {y}_2, {y}_3, {y}_4\right) \\
& =\frac{1}{4}\left[{y}_1+\operatorname{Rev}\left({y}_3\right)+\pi_{sp}^{-1}\left({y}_2+\operatorname{Rev}\left({y}_4\right)\right)\right]
\end{aligned}
\label{Eq::SCSR_Merge}
\end{equation}
\rchg{The fused sequence $z_{sq}$ is then fed into SS1D for subsequent feature processing.}

\subsection{Inference Flowchart}
\begin{algorithm}[!ht]
  \caption{SCI-Mamba Network Flowchart (Inference)}
  \label{alg:sci_mamba_flow}
  \begin{algorithmic}[1]
    \Require Low-light input image $X_0 \in \mathbb{R}^{H\times W\times 3}$; pretrained network parameters
    \Ensure Enhanced image $X_3 \in \mathbb{R}^{H\times W\times 3}$
    \State Generate $X_0^l$ via long-side constrained proportional resizing \Comment{\rchg{Long side capped at 512 px, aspect ratio preserved}}
    \State \textbf{Step 1: Patch Embedding \& Unique 2D-to-1D Conversion}
    \State $X_{0,1} \gets \operatorname{PatchEmbed}(X_0^l)$ \Comment{Downsampled feature grid $h\times w \times C$}
    \State $x_{0,1} \gets \operatorname{Flatten}(X_{0,1})$ \Comment{Only one global flatten operation}
    \State \textbf{Step 2: Stage 1 -- Initial Illumination Estimation}
    \State $z_1 \gets \operatorname{VSS1D}(x_{0,1})$ \Comment{Global multi-direction context extraction}
    \State $c_1 \gets \sigma\big(\operatorname{Linear}(z_1)\big)$ \Comment{Pixel-wise illumination compensation}
    \State $i_1 \gets x_{0,1} + c_1$ \Comment{Coarse illumination sequence}
    \State $\hat{x}_1 \gets x_{0,1} \odot i_1^{\circ(-1)}$ \Comment{1D Retinex enhancement, ~\eqref{Eq::Retinex_Sequence}}
    \State $x_1 \gets \operatorname{Rectify}(\hat{x}_1)$
    \State \textbf{Step 3: Stage 2 -- Self-calibrated Illumination Refinement}
    \State $i_{2,0} \gets \operatorname{VSS1D}(x_1)$ \Comment{First stacked VSS1D unit}
    \State $i_{2,1} \gets \operatorname{VSS1D}(i_{2,0})$ \Comment{Second stacked VSS1D unit}
    \State $i_2 \gets \operatorname{Rectify}\big(i_1 + i_{2,0} + i_{2,1}\big)$ \Comment{Refined illumination sequence}
    \State $\hat{x}_2 \gets x_{0,1} \odot i_2^{\circ(-1)}$ \Comment{Retinex based on raw base sequence}
    \State $x_2 \gets \operatorname{Rectify}(\hat{x}_2)$
    \State \textbf{Step 4: Stage 3 -- Reconstruction Branch}
    \State $i_3 \gets \operatorname{VSS1D}(x_2)$ \Comment{Final 1D illumination sequence}
    \State $I_3 \gets \operatorname{Reshape}(i_3)$ \Comment{Only one global 1D-to-2D recovery}
    \State $I_3 \gets \operatorname{UpSample}(I_3)$ \Comment{Recover full input resolution}
    \State $\hat{X}_3 \gets X_0 \odot I_3^{\circ(-1)}$ \Comment{Image-domain Retinex \rchg{on original image}, ~\eqref{Eq::Retinex_Image}}
    \State $X_3 \gets \operatorname{Conv2D}\big(\operatorname{Rectify}(\hat{X}_3)\big)$ \Comment{Final refinement convolution}
    \State \Return $X_3$
  \end{algorithmic}
\end{algorithm}

We summarize the complete \rchg{inference} procedure of SCI-Mamba in Algorithm~\ref{alg:sci_mamba_flow}.
\rchg{The algorithm provides the overall execution flow, and the following description focuses on the inference-time downsampling strategy for high-resolution inputs.}

\rchg{To reduce memory and computation for ultra-high-resolution inputs, we adopt long-side constrained resizing: if the longer dimension of input $X_0$ exceeds the preset threshold (512 pixels), we proportionally shrink the image while preserving the aspect ratio; otherwise the raw resolution is retained. The network predicts low-resolution illumination maps on the scaled input, then upsamples them to the original dimensions for Retinex decomposition on the unmodified high-resolution image. This balances inference latency and texture preservation.}

\section{Loss Functions} \label{Sec::Loss_Function}
We construct a five-term weighted multi-objective total loss to jointly constrain color distortion, spatial smoothness, RGB channel balance, structural fidelity and targeted differential enhancement for space foreground/background. The unified total loss formulation is
\begin{equation}
\begin{aligned}
\operatorname{Loss}_{total}=\sum_{k=1}^{5}\alpha_k \operatorname{Loss}_{k},
\label{Eq::Stage_Loss}
\end{aligned}
\end{equation}
where $\alpha_k$ denotes the weight coefficient balancing the contribution of each sub-loss during training. Each loss term targets distinct degradation artifacts in orbital low-light imagery, with clear functional differentiation detailed separately below.

\subsection{Illumination Color Constancy Loss}
The illumination color constancy loss ($\operatorname{Loss}_1$) constrains RGB channel balance on the intermediate illumination sequences $i_1,i_2,i_3$ generated by the three stages. Physically, illumination maps represent scene light intensity distribution and should not carry inherent color bias. This term penalizes channel mean divergence of each stage's illumination sequence to suppress color shift introduced during sequence-domain Retinex decomposition.
\begin{equation}
\begin{aligned}
\operatorname{Loss}_{1}
= & \sum_{s=1}^{3}\beta_{1,s}\left[
(\mu_{s,r}-\mu_{s,g})^2\right.\\
&\left.+(\mu_{s,r}-\mu_{s,b})^2
+(\mu_{s,g}-\mu_{s,b})^2\right]^{1/2}
\end{aligned}
\label{Eq::Loss1_Color}
\end{equation}
$\beta_{1,s}$ denotes stage-specific weight; $\mu_{s,r},\mu_{s,g},\mu_{s,b}$ are channel-wise mean values of stage-$s$ illumination sequence $i_s$. Note that this loss operates exclusively on illumination intermediate features, differing fundamentally from $\operatorname{Loss}_3$ which regularizes the final enhanced output image's channel balance.

\subsection{Smoothness Loss}
The smoothness loss ($\operatorname{Loss}_2$) \cite{ma2025learning} enforces natural spatial continuity on the final enhanced image $X_3$ with structure-adaptive weighting derived from raw input $X_0$. For flat, textureless deep-space background regions, strong smooth regularization suppresses amplified sensor noise; for spacecraft edges and structural textures, regularization weight decays automatically to avoid over-blurring critical detail
\begin{equation}
\begin{aligned}
\operatorname{Loss}_{2}= \frac{1}{hw}&\sum_{i,j}\sum_{(k,l)\in D_{i,j}}
 w_{k,l}(i,j)\times \\
& \left\|X_3(i+k,j+l)-X_3(i,j)\right\|_{1}
\end{aligned}
\label{Eq::Loss2_Smoothness}
\end{equation}
where $h,w$ represent output image height and width; $D_{i,j}$ defines pixel offset set within a $(2\zeta+1)\times(2\zeta+1)$ local neighborhood excluding center coordinate as below
\begin{equation}
\begin{aligned}
{D}_{i,j}=\big\{(k,l)\ \big|\ &k,l\in[-\zeta,\zeta]\cap\mathbb{Z},\ (k,l)\neq(0,0),\\
&1\leq i+k\leq h,\ 1\leq j+l\leq w\big\}
\end{aligned}
\label{Eq::SmoothNeighborSet}
\end{equation}
Structure-aware weight $w_{k,l}(i,j)$ is calculated in YCrCb color space from input local color differences
\begin{equation}
\begin{aligned}
w_{k,l}(i,j) = \exp\Big( -\frac{1}{2\sigma_s^2}&\sum_{c\in\{Y,C_r,C_b\}}
 \left[ C_{0,c}(i+k,j+l) \right. \\
& \left. - C_{0,c}(i,j) \right]^2 \Big)
\end{aligned}
\label{Eq::SmoothWeight}
\end{equation}
When adjacent pixels share nearly identical YCrCb values (flat background), $w_{k,l}(i,j)\to1$ to strengthen smooth penalty; large color gradients at edges reduce weight to preserve structural sharpness.

\rchg{
$X_3(i,j)\in\mathbb{R}^{3}$ denotes the RGB pixel vector at spatial location $(i,j)$ of the enhanced image:
$X_3(i,j)=\left[R(i,j),\,G(i,j),\,B(i,j)\right]^{\mathrm T}$.
The operator $\|\cdot\|_1$ denotes the $\ell_1$ norm over the RGB channel dimension. Therefore,
\begin{equation}
\begin{aligned}
&\left\|X_3(i+k,j+l)-X_3(i,j)\right\|_1 \\
={}&
\left|R(i+k,j+l)-R(i,j)\right| \\
&+
\left|G(i+k,j+l)-G(i,j)\right| \\
&+
\left|B(i+k,j+l)-B(i,j)\right|
\end{aligned}
\label{Eq::RGB_L1_Difference}
\end{equation}
$\left\|X_3(i+k,j+l)-X_3(i,j)\right\|_1$
represents the sum of the absolute differences of the three RGB channels between the neighboring pixel at $(i+k,j+l)$ and the center pixel at $(i,j)$.
}

\subsection{RGB Relative Response Constraint Loss}
The RGB relative response constraint loss ($\operatorname{Loss}_3$) directly regularizes the final enhanced image $X_3$ to retain the original input's RGB proportional balance, preventing unnatural color distortion while allowing reasonable overall brightness adjustment. Distinct from $\operatorname{Loss}_1$ (constraining illumination maps), this loss targets pixel-level channel ratios of the visible output
\begin{equation}
\operatorname{Loss}_{3}= \frac{1}{3h\cdot w}\sum_{p}\sum_{c\in\{r,g,b\}}\left|R_{X_3,c}(p)-R_{X_0,c}(p)\right|
\label{Eq::Loss3_Ratio_Align}
\end{equation}
Per-pixel relative channel response for arbitrary image $X$ is defined as
\begin{equation}
R_{X,c}(p)=\frac{X_c(p)}{\bar{X}(p)+\varepsilon_{3}}
\label{Eq::RelativeRGBResponse}
\end{equation}
$\bar{X}(p)$ denotes average RGB intensity at pixel $p$, and tiny constant $\varepsilon_3=1\times10^{-3}$ prevents division-by-zero numerical overflow.

\subsection{Fidelity Loss }
The fidelity loss ($\operatorname{Loss}_4$) \cite{ma2025learning} restricts structural deviation between each stage's intermediate enhanced sequence and the raw low-light input, eliminating extreme over-enhancement or geometric distortion in the absence of paired well-exposed supervision.
\begin{equation}
\operatorname{Loss}_{4}=\frac{1}{3h\cdot w}\sum_{s=1}^{3}\sum_{p}\sum_{c\in\{r,g,b\}}\beta_{4,s}\left(X_{s,c}(p)-X_{0,c}(p)\right)^2
\label{Eq::Loss4_Fidelity}
\end{equation}
$\beta_{4,s}$ assigns progressive weight to each stage's intermediate output $X_s$, penalizing large pixel-wise divergence from raw input $X_0$.

\subsection{Prior-guided Illumination Monitor}
The prior-guided illumination monitor ($\operatorname{Loss}_5$) is \rchg{introduced to provide region-specific illumination guidance for foreground enhancement and background suppression}.
It dynamically generates a target illumination guidance map $I_{X_0}$ purely from input brightness and gradient features without manual labeling: texture-rich spacecraft regions receive low target illumination values (strong Retinex enhancement), while smooth dark cosmic backgrounds are assigned high target illumination to suppress noise amplification. The loss computes mean absolute error between each stage's single-channel illumination map and the auto-generated guidance map.
\begin{equation}
\operatorname{Loss}_{5}=\frac{1}{h\cdot w}\sum_{s=1}^{3}\sum_{p}\beta_{5,s}\left|Gray(I_s(p))-I_{X_0}(p)\right|
\label{Eq::Loss5}
\end{equation}
where $\beta_{5,s}$ is the stage-wise weight for the fifth loss term. Grayscale conversion formula for RGB illumination map $P$ follows standard luminance weights.
\begin{equation}
Gray(P)=0.299P_{R}+0.587P_{G}+0.114P_{B}
\label{Eq::PreprocessedGray}
\end{equation}
The target illumination guidance map $I_{X_0}$ integrates two self-extracted masks: bright-region map $X_{\text{bright}}$ and texture feature map $X_{\text{texture}}$, with separate weights for dark-texture and bright-texture spacecraft areas.
\begin{equation}
\begin{aligned}
I_{X_0}=1- [\lambda_d (1-X_{\text{bright}})X_{\text{texture}}+\lambda_bX_{\text{bright}}X_{\text{texture}} ]
\end{aligned}
\label{Eq::TargetIllu}
\end{equation}
$\lambda_d=1.35$ (dark texture weight), $\lambda_b=0.12$ (bright texture weight); term $(1-X_{\text{bright}})X_{\text{texture}}$ marks shadowed spacecraft structures, while $X_{\text{bright}}X_{\text{texture}}$ labels illuminated satellite surfaces. The construction pipelines for $X_{\text{bright}}$ and $X_{\text{texture}}$ are detailed sequentially below.

\subsubsection{$X_{\text{bright}}$ Computation}
Two prior guidance maps $X_\mathrm{bright}$ and $X_\mathrm{texture}$ are constructed to weight the multi-objective loss, which are derived from luminance and gradient features respectively.
The bright-region guidance map segments high-brightness pixels from smoothed input luminance.
\begin{equation}
X_{\text{bright}}(p)=G_{\sigma_b}\left(\cfrac{\widetilde{Y}(p)-T_b}{P_b-T_b}\right)
\label{Eq::BrightMask}
\end{equation}
$\widetilde{Y}$ is denoised luminance after grayscale conversion, median filtering and Gaussian smoothing.
\begin{equation}
\widetilde{Y}=G_{\sigma_y}\left(\operatorname{Median}\left(Gray({X_0}(p))\right)\right)
\label{Eq::SmoothedLuminance}
\end{equation}
\rchg{Otsu \cite{otsu1975threshold} adaptive threshold scaled by coefficient $\alpha_b$ sets brightness segmentation boundary.}
\begin{equation}
T_b=\alpha_b\operatorname{Otsu}\left(\widetilde{Y}\right)
\label{Eq::BrightnessThreshold}
\end{equation}
$P_b=99.5\%$ denotes the 99.5th percentile luminance value as upper softening bound for bright pixels.

\subsubsection{$X_{\text{texture}}$ Computation}
The texture guidance map quantifies local structural confidence via gradient magnitude and morphological segmentation.
\begin{equation}
X_{\text{texture}}(p)=\left[M_s(p)\left(\theta_0+\theta_1E_s(p)\right)\right]
\label{Eq::TextureGuidance}
\end{equation}
$\theta_0=0.45$ \rchg{denotes the baseline structural confidence, $\theta_1=0.55$ controls the contribution of the gradient response}; $M_s$ is smoothed soft region mask derived from morphological processed edge seed map $E_0$.
\begin{equation}
M_s=G_{\sigma_f}\left(M_f\right)
\label{Eq::SoftRegionSupport}
\end{equation}
Hard structural mask $M_f$ is generated via cascaded morphological operations on raw edge seeds.
\begin{equation}
\begin{aligned}
M_f = \operatorname{Close}\Big( \operatorname{AreaOpen}\big( \operatorname{Dilate}\big( \operatorname{Fill}\big( \\
\qquad \operatorname{Close}\big( \operatorname{Dilate}(E_0) \big) \big) \big), A_{\min} \big) \Big)
\end{aligned}
\label{Eq::FeatureRegionMask}
\end{equation}
\rchg{The binary structural mask is refined through a sequence of morphological operations. Dilation and closing are first applied to connect adjacent structural responses, followed by hole filling to obtain continuous regions. Area opening is then used to remove isolated regions with an area smaller than $A_{\min}=300$, and a final closing operation is applied to further improve region connectivity.}\
Edge seed map $E_0$ binarizes normalized gradient magnitude via Otsu thresholding.
\begin{equation}
E_0(p)=\left\{
\begin{array}{rcl}
1, &\textit{if} & \widehat{G}_{otsu}(p)\geq \alpha_g,\\
0, &\textit{if} & \widehat{G}_{otsu}(p)< \alpha_g
\end{array}\right.
\label{Eq::EdgeSeed}
\end{equation}
$\widehat{G}_{otsu}=\operatorname{Otsu}(\widehat{G})$, gradient threshold coefficient $\alpha_g=0.82$. Normalized gradient magnitude.
\begin{equation}
\widehat{G}=\cfrac{G-\min(G)}{\max(G)-\min(G)}
\label{Eq::GradientNormalize}
\end{equation}
Raw gradient magnitude from smoothed luminance $\widetilde{Y}$.
\begin{equation}
G=\sqrt{\left(\nabla_x\widetilde{Y}\right)^2+\left(\nabla_y\widetilde{Y}\right)^2}
\label{Eq::GradientMagnitude}
\end{equation}
Soft edge confidence map $E_s$ smooths normalized gradient within structural mask boundaries.
\begin{equation}
E_s(p)=G_{\sigma_e}\left[\left(\frac{\widehat{G}(p)-P_l}{P_h-P_l+\epsilon}\right)M_f(p)\right],
\label{Eq::SoftEdgeConfidence}
\end{equation}
$P_l=85\%, P_h=99\%$ are lower/upper gradient percentiles, $\sigma_e=0.6$ Gaussian smooth width.

\rchg{All guidance masks ($X_{\mathrm{bright}}$, $X_{\mathrm{texture}}$, and the resulting $I_{X_0}$) are computed offline before training following Eq.~\eqref{Eq::BrightMask}--\eqref{Eq::SoftEdgeConfidence}; no additional learnable parameters are introduced.}

\section{Dataset and Training} \label{Sec::Space_Dark_and_Training}

\subsection{Space Dark-1.0 Dataset}\label{SubSec::Dataset}
Existing orbital target datasets are optimized for pose estimation and segmentation tasks and lack authentic low-light orbital measurements, as summarized in Section~\ref{Sec::Introduction}. To resolve this data shortage for space low-light enhancement research, we construct the hybrid benchmark \textit{Space Dark-1.0} integrating three complementary data sources that jointly mitigate synthetic-to-real domain gaps.
\rchg{The dataset comprises three complementary components: physically constrained synthetic spacecraft images generated using PASim, an in-house simulation platform developed on Unreal Engine, with orbital dynamics, attitude motion, space illumination, and surface reflectance taken into account (4,700 images); ground-based darkroom images of scaled spacecraft models acquired under controlled low-light conditions using a real imaging system, providing realistic sensor responses and illumination variations (2,600 images); and real on-orbit images of the H-IIA upper stage extracted from publicly released imagery of Astroscale's ADRAS-J mission, representing authentic spacecraft observations under different viewing and illumination conditions (2,100 images).}
Further details on the dataset construction and composition are provided in Section I \rchg{of the supplementary appendix}.

The full dataset totals 9,400 images, forming \rchg{a dedicated benchmark covering} synthetic simulation, semi-physical ground verification and real orbital operating environments.\rchg{The training and test sets are strictly non-overlapping, split at the individual-image level, without temporally correlated frames.}

\subsection{Training Process}
Orbital satellite imagery exhibits limited color diversity, which may bias color correction modules during training. We supplement our primary training set with external public low-light datasets for balanced color distribution: 1,600 randomly sampled frames from ExDark \cite{loh2019getting}, 8,000 samples from \textit{Space Dark-1.0}, and 2,400 images from SPEED+ \cite{park2022speed+}. The model is trained under the five-term joint loss defined in \eqref{Eq::Stage_Loss}; loss computation relies on intermediate stage sequences during training, while inference strips auxiliary loss branches and retains only the lightweight all-sequence forward pipeline for speed.
\rchg{\
To ensure a fair comparison, both training and inference evaluation were conducted on the same NVIDIA Tesla V100-PCIe-32GB GPU platform under consistent runtime configurations.
}\looseness=-1

The patch embedding module adopts a patch size of 4, generating feature grids with dimension $H/4 \times W/4$. For the VSS1D module formulated in \eqref{Eq::VSS1D}, the input sequence channel dimension \rchg{$C=64$} and Mamba hidden state dimension \rchg{$N=16$}. The 1D convolution kernel in \eqref{Eq::SS1D_Conv} is set to $1\times3$, and the linear expansion factor for feature splitting in  \eqref{Eq::SS1D_Split} equals 2. The time step clipping bounds in  \eqref{Eq::Mamba_Delta} are assigned as $\Delta_{\min}=1\times10^{-5}, \Delta_{\max}=0.1$.

For the total joint loss  \eqref{Eq::Stage_Loss}, the balancing coefficients are set as $\alpha_1=0.006$, $\alpha_2=0.003$, $\alpha_3=0.25$, $\alpha_4=0.006$, $\alpha_5=2.2$.
Stage-wise weights $\beta_{s,1}=0.50,\beta_{s,2}=0.75,\beta_{s,3}=1.00$ are shared by the color constancy \eqref{Eq::Loss1_Color}, fidelity \eqref{Eq::Loss4_Fidelity}, and illumination monitor \eqref{Eq::Loss5} losses.
The smoothness loss  \eqref{Eq::Loss2_Smoothness} uses neighborhood radius $\zeta=2$ and Gaussian smoothing coefficient $\sigma_s=10$.
A tiny constant $\varepsilon_3=1\times10^{-3}$ is adopted to avoid division overflow in RGB relative response loss \eqref{Eq::RelativeRGBResponse}.
Coefficients for the illumination guidance map construction in \eqref{Eq::TargetIllu} -- \eqref{Eq::TextureGuidance} are set as follows:
$\lambda_d=1.35,\lambda_b=0.12,\alpha_b=0.78,\sigma_b=0.8,\sigma_y=0.6,P_b=99.5\%,\theta_0=0.45,\theta_1=0.55,\sigma_f=1.2,A_{\min}=300,\alpha_g=0.82,\sigma_e=0.6$.
The generated target illumination map $I_{X_0}$ is clamped within the range $[0.1,1.0]$, with gradient percentile thresholds $P_l=85\%,P_h=99\%$.

We adopt the AdamW optimizer with initial learning rate $1\times10^{-4}$ and weight decay $1\times10^{-4}$. The batch size is 4 with 2-step gradient accumulation, and the training process runs for a total of 500 epochs. The learning rate schedule includes a linear warm-up phase within the first 5 epochs (warm-up ratio from 0.2 to 1.0), followed by cosine annealing that decays the learning rate down to $5\times10^{-6}$. Gradient clipping with maximum norm 1.0 is applied to stabilize training, and the global random seed is fixed to 2 for full experimental reproducibility.

Table~\ref{Table::TrainingResult} compares training overhead metrics among SCI-Mamba and three representative unsupervised enhancement methods. \rchg{All methods were trained on the same 12,000-image training set and platform; each baseline followed its default optimization protocol.}
\begin{table*}[htpb]
  \centering
\begin{footnotesize}
  \caption{Training Overhead Comparison Across Unsupervised Enhancement Algorithms}\label{Table::TrainingResult}
\begin{tabular}{ c|c|c|c  }
  \hline \hline
  \textbf{Metrics} & Training Time (s) & Peak Memory (MB) & Max./Avg. GPU Utilization   \\ \hline
  SCI++       & 337915.12 & 1593.02 & 65.00\%, 24.37\% \\
  Zero-DCE++    & 95240.62  & 1974.31 & 92.00\%, 29.76\% \\
  RUAS        & 747616.60  & 415.59  & 39.00\%, 11.63\% \\ \hline
  SCI-Mamba (ours) & 371553.58 & 1657.21 & 60.00\%, 24.99\% \\
  \hline \hline
\end{tabular}
\end{footnotesize}
\end{table*}
\rchg{SCI-Mamba's training time (371,553 s) sits between the lightweight Zero-DCE++ (95,240 s) and the slower RUAS (747,617 s), being moderately higher than SCI++ (337,915 s). This overhead mainly reflects the 500-epoch schedule over 12,000 samples and the five-term joint loss supervising all three stages; since these auxiliary branches are removed at inference, it does not affect the real-time deployment performance reported below.}

\section{Inference Experiment} \label{Sec::Experiment}
We conduct comparative experiments covering three mainstream algorithm families: CNN, Transformer and Mamba. The competing methods include SCI++ \cite{ma2025learning}, Zero-DCE++ \cite{li2021learning}, RUAS \cite{liu2021retinex}, \mbox{ECMamba} \cite{dong2024ecmamba}, WalMaFa \cite{tan2024wavelet}, LLFlow \cite{wang2022low}, Uformer \cite{wang2022uformer}, \mbox{LLFormer} \cite{wang2023ultra}, and UHDformer \cite{wang2024correlation}.
\rchg{\
The test set includes 1,400 unseen disjoint samples from the \textit{Space Dark-1.0} dataset and 100 images of previously unseen satellite types from the SPARK 2021 dataset \cite{musallam2021spacecraft}.
}\
\rchg{All models run on identical hardware. Multiple warm-up passes precede latency recording to exclude CUDA initialization and cache preloading; the measured frame rate covers the full inference pipeline including resizing, padding and model forward inference, while image storage and debug logs are excluded, reflecting real on-orbit processing latency.}
\rchg{\
SCI++, Zero-DCE++, and RUAS were evaluated using weights trained on the same 12,000-image training set, which includes 8,000 images from Space Dark-1.0, whereas ECMamba, WalMaFa, LLFlow, Uformer, LLFormer, and UHDformer were evaluated using their publicly available pretrained weights. The former three are unsupervised and can therefore be retrained on our unpaired low-light training data, whereas the latter six are supervised methods that require paired low-light and normally exposed training images, which are not available for orbital low-light scenarios; they are thus evaluated with their publicly released pretrained weights and without any domain adaptation, keeping the comparison consistent and fair.
}

\rchg{To verify the efficiency gain of the single-flattening design, we build an ablation variant SCI-Mamba(2D).}\
It shares identical loss functions, training strategies and three-stage illumination refinement pipeline with the proposed SCI-Mamba, while replacing our dedicated 1D-only VSS1D module with the original layer-wise 2D SS2D scanning from VMamba \cite{liu2024vmamba}. We adopt four quantitative metrics to evaluate model deployment potential: average inference FPS, peak GPU memory consumption, theoretical GFLOPs and total trainable parameters (in millions). The parameter count is defined as $\operatorname{Params}=\sum_{\theta_t\in\Theta}\operatorname{numel}(\theta_t)$, where $\Theta$ represents the collection of all trainable weight tensors.

\begin{figure*}[htbp]
	\centering \includegraphics[width=0.9\linewidth]{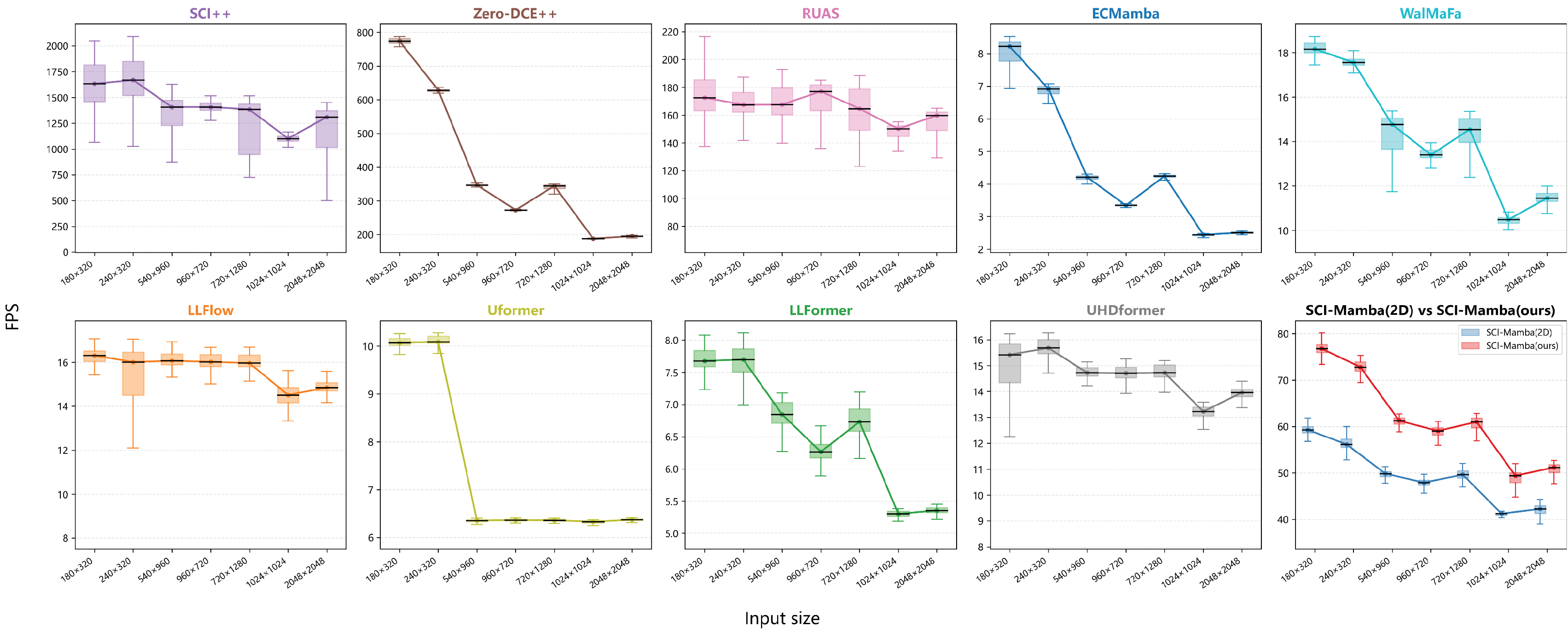}
	\caption{Inference Frame Rate Boxplot for Multi-Resolution Test Pictures}
	\label{Fig::FrameRates}
\end{figure*}

\rchg{Figure~\ref{Fig::FrameRates} presents the inference-speed distributions across multi-resolution test images. Lightweight CNN-based methods achieve the highest frame rates, while the evaluated Transformer- and Mamba-based methods generally require greater computational resources. Under the same GPU setting, SCI-Mamba runs faster than the evaluated Transformer- and Mamba-based baselines. Detailed quantitative results are reported in Table~\ref{Table::TestResult}.}

\begin{table*}[htpb]
\centering
\begin{footnotesize}
  \caption{Quantitative Inference Performance Comparison}\label{Table::TestResult}
\begin{tabular}{ c|c|c|c|c }
    \hline \hline
  \textbf{Methods} & Avg. FPS & Peak Mem. (MB) & GFLOPs & Params (M) \\ \hline
  SCI++      & 1361.18  & 225.00   & 0.57    & \(2.29\times10^{-2}\) \\
  Zero-DCE++ & 196.44  & 3683.59  & 184.42  & \(7.92\times10^{-2}\) \\
  RUAS       & 161.97  & 273.63   & 1.71    & \(3.44\times10^{-3}\) \\
  ECMamba    & 2.54    & 19232.76 & 184.04  & 1.99 \\
  WalMaFa    & 11.84   & 4304.04  & 482.56  & 11.63 \\
  LLFlow     & 15.15   & 2392.08  & 112.86  & 5.43 \\
  Uformer    & 6.38    & 8768.03  & 715.64  & 50.88 \\
  LLFormer   & 5.42    & 1027.09  & 1166.94 & 24.55 \\
  UHDformer  & 14.15   & 2223.67  & 12.09   & \(3.39\times10^{-1}\) \\ \hline
  SCI-Mamba(2D)  & 43.47 & 763.57 & 3.17  & \(4.55\times10^{-1}\) \\
  SCI-Mamba(ours) & 52.14 & 394.58 & 4.74  & \(3.85\times10^{-1}\) \\
  \hline \hline
\end{tabular}
\end{footnotesize}
\end{table*}

Ablation comparison between SCI-Mamba and its 2D baseline \rchg{confirms} the efficiency improvement of our all-sequence architecture: the average inference FPS rises from \rchg{43.47 to 52.14, delivering a 19.9\%} speedup.
\rchg{This acceleration results from the single-flattening design analyzed in Section \ref{Sec::MainResult}, which removes the layer-wise conversion bottleneck of conventional VMamba pipelines.}\
\rchg{Although lightweight CNN models are faster, their poor restoration quality restricts on-orbit deployment, while Mamba/Transformer competitors face prohibitive memory requirements: ECMamba peaks at 19,232 MB, WalMaFa at 4,304 MB, and all Transformer variants occupy gigabyte-level video memory, well above satellite embedded budgets.}

\rchg{In terms of computational cost and model scale, SCI-Mamba incurs only 1/38.8 of ECMamba's GFLOPs and 1/101.8 of WalMaFa's GFLOPs; its parameter count is merely 1/5.2 and 1/30.2 of the two mainstream Mamba competitors, respectively. For inference speed, our method runs $20.53\times$ faster than ECMamba and $4.40\times$ faster than WalMaFa. Compared with Transformer- and flow-based baselines LLFlow, Uformer, LLFormer and UHDformer, our speed gains range from $3.44\times$ to $9.62\times$, achieving real-time inference speed that attention-based models cannot reach.}
\rchg{SCI-Mamba's peak GPU memory is 394.58 MB, lower than all evaluated Mamba/Transformer competitors and within the budget of low-power spaceborne hardware.}

\rchg{
Since paired well-exposed reference images are unavailable for the low-light orbital imagery, we employ five no-reference image quality assessment metrics to quantitatively evaluate the enhancement results: NIQE \cite{mittal2012making}, BRISQUE \cite{mittal2012no}, PIQE \cite{venkatanath2015blind}, ARNIQA \cite{agnolucci2024arniqa}, and MANIQA \cite{yang2022maniqa}. In addition to conventional metrics based on natural image statistics and local distortion analysis, ARNIQA and MANIQA provide complementary perceptual quality assessments through learned distortion representations and multidimensional attention features, respectively.
Lower NIQE, BRISQUE, and PIQE values indicate better quality, whereas higher ARNIQA and MANIQA values are better. The metrics are described as follows: NIQE evaluates consistency with natural image statistics to measure overall scene authenticity; BRISQUE quantifies structural damage, noise amplification, and over-enhancement artifacts from spatial statistics; and PIQE detects local block distortion to reflect micro-perceptual quality.}
\rchg{\
ARNIQA learns a representation of the image distortion manifold through self-supervised contrastive learning and predicts perceptual quality scores using a linear regressor. MANIQA models global and local relationships among image features through attention mechanisms across channel and spatial dimensions, aggregating patch-level quality predictions with learned weights to obtain an overall perceptual quality score.
}

\begin{table*}[htpb]
  \centering
  \footnotesize
  \caption{No-Reference Image Quality Assessment Metrics}
  \label{Table::TestResult2}
  \begin{tabular}{ c|c|c|c|c|c }
  \hline \hline
    \textbf{Methods} & NIQE & BRISQUE & PIQE & ARNIQA  & MANIQA \\ \hline
    Raw Input        & 19.98 & 54.69 & 54.05 & \(4.96\times10^{-1}\)   & \(2.84\times10^{-1}\)\\ \hline
    SCI++        & 21.72 & 64.16 & 70.19 & \(4.53\times10^{-1}\)  & \(2.72\times10^{-1}\)  \\
    Zero-DCE++   & 24.36 & 56.06 & 57.52 & \(4.14\times10^{-1}\)  & \(2.57\times10^{-1}\)  \\
    RUAS         & 24.28 & 61.39 & 70.95 & \(4.19\times10^{-1}\)  & \(2.63\times10^{-1}\)  \\
    ECMamba      & 21.42 & 51.62 & 72.19 & \(4.20\times10^{-1}\)  & \(2.62\times10^{-1}\)  \\
    WalMaFa      & 21.49 & 68.08 & 75.57 & \(4.53\times10^{-1}\)  & \(2.84\times10^{-1}\)  \\
    LLFlow       & 24.45 & 62.97 & 76.63 & \(4.42\times10^{-1}\)  & \(2.66\times10^{-1}\)  \\
    Uformer      & 22.06 & 91.69 & 86.43 & \(4.77\times10^{-1}\)  & \(2.75\times10^{-1}\)  \\
    LLFormer     & 20.07 & 65.86 & 69.27 & \(4.35\times10^{-1}\)  & \(2.64\times10^{-1}\)  \\
    UHDformer    & 21.22 & 69.68 & 75.92 & \(4.26\times10^{-1}\)  & \(2.59\times10^{-1}\)  \\ \hline
    SCI-Mamba(ours) & 20.15 & 56.54 & 54.21 & \(4.85\times10^{-1}\)  & \(2.79\times10^{-1}\)  \\
    \hline \hline
  \end{tabular}
\end{table*}

From Table~\ref{Table::TestResult2}, we analyze the visual performance of SCI-Mamba against all peers. NIQE: SCI-Mamba achieves a NIQE score of 20.15, ranking second only to LLFormer (20.07) among all competing methods, while \rchg{outperforming} all CNN-based and most Transformer/Mamba baselines. This indicates that our enhancement introduces minimal unnatural brightness or color bias to orbital imagery. BRISQUE: Although ECMamba achieves the lowest BRISQUE score (\rchg{51.62}), its 19 GB memory footprint precludes satellite deployment. Zero-DCE++ attains a comparable BRISQUE of \rchg{56.06}, yet produces severe visual artifacts as demonstrated in the qualitative results. SCI-Mamba's BRISQUE of \rchg{56.54} is a competitive value given its compact memory footprint. PIQE: Our result (\rchg{54.21}) is marginally higher than the unprocessed input (\rchg{54.05}), yet it \rchg{remains lower than all other enhancement baselines}. This \rchg{indicates} that SCI-Mamba avoids block artifacts during targeted foreground enhancement for spacecraft targets.
\rchg{\
ARNIQA: SCI-Mamba achieves the highest score among all evaluated enhancement methods, with $4.85\times10^{-1}$ compared with $4.77\times10^{-1}$ for the second-ranked Uformer. Its score is also close to that of the raw input ($4.96\times10^{-1}$). These results indicate that SCI-Mamba produces enhanced images with better perceptual quality than the other enhancement methods as measured by ARNIQA.
MANIQA: SCI-Mamba obtains a score of $2.79\times10^{-1}$, ranking second among all evaluated enhancement methods, behind \mbox{WalMaFa} ($2.84\times10^{-1}$). Meanwhile, SCI-Mamba achieves an average inference speed of 52.14 FPS compared with 11.84 FPS for WalMaFa, while requiring only 394.58 MB of peak GPU memory compared with 4304.04 MB. This quality comes at far lower computational cost than the Mamba/Transformer baselines.
}

Considering both inference efficiency and the five no-reference metrics, SCI-Mamba attains the best PIQE and ARNIQA and the second-best NIQE and MANIQA among the evaluated enhancement methods, while maintaining high inference speed and low GPU memory. Its ARNIQA and MANIQA remain slightly below the raw input; this reflects the limitation of no-reference metrics under low light rather than lower enhancement quality, since these metrics assess deviations from natural-image statistics or learned perceptual models rather than the recovery of faint spacecraft structures. Improvements in spacecraft foreground visibility are therefore assessed through qualitative comparisons and downstream detection experiments.

The supplementary appendix covers \textit{Space Dark-1.0} construction (I), loss-term ablation (II), qualitative comparisons (III), edge inference on Jetson Orin NX (IV), downstream spacecraft detection (V), and per-subset no-reference metrics for all compared methods (VI).

\section{Conclusion} \label{Sec::Conclusion}
We propose SCI-Mamba, an unsupervised low-light enhancement network tailored for non-cooperative orbital targets.
\rchg{By integrating SCI++'s self-calibration, VMamba's linear-complexity state-space modeling, and Retinex priors, this fully sequence-dominated pipeline performs a single global flattening at input and a single reconstruction at output, avoiding the per-layer 2D/1D conversion overhead of conventional restoration architectures.}
The three-stage progressive refinement workflow is regularized by a multi-term weighted loss function, where a prior-guided illumination monitor adaptively weights enhancement strength for textured spacecraft foreground and smooth dark background. To address the scarcity of low-light orbital data, we build the hybrid benchmark \textit{Space Dark-1.0}, which aggregates synthetic, ground hardware-in-the-loop and real on-orbit samples to support unsupervised training.
\rchg{Ablation and comparative experiments on synthetic, darkroom and on-orbit scenes show that SCI-Mamba preserves spacecraft structure while suppressing background noise and, through its sequence-dominated design, achieves markedly higher inference speed than the evaluated baselines.}

\rchg{Future work will focus on lightweight backbones for spaceborne deployment,  video enhancement,  downstream pose estimation and detection.}

\bibliographystyle{ieeetr}
\bibliography{Ref}

\begin{IEEEbiography}[{\includegraphics[width=1in,height=1.25in,clip,keepaspectratio]{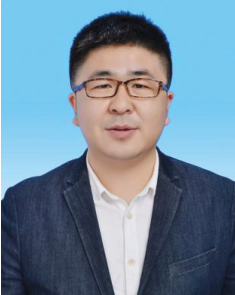}}]{Yiyong Sun}
received the Ph.D.  degree in control science and engineering from Harbin Institute of Technology, Harbin, China,  2017.  He is currently an Assistant Professor with  the School of Aerospace Engineering, Beijing Institute of Technology, Beijing, China. Before June 2021, he worked as a Research Assistant and a Postdoctoral Fellow with the Department of Automation, Tsinghua University, Beijing. He ever was a Research Assistant with the Chair of Automatic Control Engineering (LSR), Technical University of Munich, Munich, Germany, from 2015 to 2016. His research interests include advanced observer, robotics and adaptive learning control.
\end{IEEEbiography}

\begin{IEEEbiography}[{\includegraphics[width=1in,height=1.25in,clip,keepaspectratio]{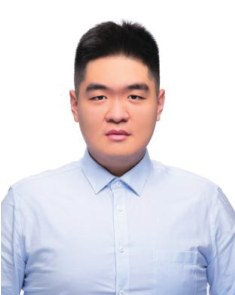}}]{Weihang Shan}
received the B.E. degree in Aerospace Engineering from Beijing Institute of Technology, Beijing, China, in 2022. He is currently working toward the M.Sc. degree in Aeronautical and Astronautical Science and Technology with the School of Aerospace Engineering, Beijing Institute of Technology, Beijing, China. His research interests include navigation, deep learning and computer vision.
\end{IEEEbiography}

\begin{IEEEbiography}[{\includegraphics[width=1in,height=1.25in,clip,keepaspectratio]{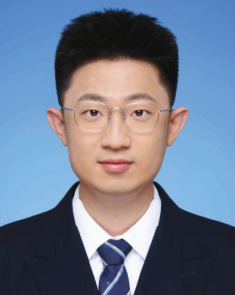}}]{Shijun Wei}
 was born in 1997. He received the B.S. degree in Flight Vehicle Design and Engineering from the School of Aerospace Engineering, Beijing Institute of Technology, Beijing, China, in 2019, where he is currently working toward the Ph.D. degree in Aeronautical and Astronautical Science and Technology.
His research interests include state estimation and maneuvering target tracking, and their applications in engineering.
\end{IEEEbiography}

\begin{IEEEbiography}[{\includegraphics[width=1in,height=1.25in,clip,keepaspectratio]{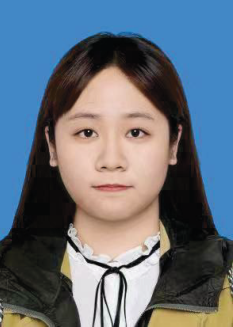}}]{Diwei Zhou}
was born in Chongqing, China, in 2002. She received the B.S. degree in aeronautical and astronautical science and technology from Beijing Institute of Technology, China, in 2024, where she is currently pursuing the M.S degree.
\end{IEEEbiography}

\begin{IEEEbiography}[{\includegraphics[width=1in,height=1.25in,clip,keepaspectratio]{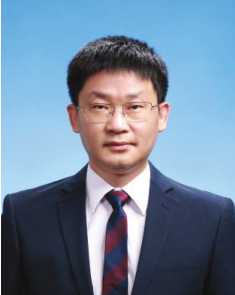}}]{Guang Zhai}
was born in Hebei, China, in 1979. He received the B.S. degree in polymer science and technology, and M.S. degree in mechanical and electronics engineering from Harbin Institute of Technology, Harbin, China, in 2001 and 2005, respectively, and the Ph.D. degree in control science and engineering from the Department of Control Science and Engineering, and Harbin Institute of Technology, Harbin, China, in 2009. He is currently working in Aerospace Engineering School, Beijing Institute of Technology, Beijing, China, as a Professor. His research interests include  dynamics,   autonomous navigation and control.
\end{IEEEbiography}

\end{document}